# Non-volatile memory and IR radiation modulators based upon graphene-on-ferroelectric substrate. A review

Strikha M. V.

V. Lashkaryov Institute of Semiconductor Physics, National Academy of Science of Ukraine, 41 Nauka Ave., 03650 Kyiv-28, Ukraine, maksym_strikha@hotmail.com

**Abstract**. I present a review of both experimental and theoretical studies performed during the recent three years, which deal with the physical properties and possible applications of graphene placed on ferroelectric (organic or $Pb(Zr_xTi_{1-x})O_3$ (PZT) ceramic) substrates. The system 'graphene-on-ferroelectric substrate' has unique advantages in comparison with the graphene deposited on $SiO_2$ or on dielectrics with high dielectric permittivity. In particular, one can obtain high ($\sim 10^{12}$ cm$^{-2}$) carrier concentrations in the doped graphene-on-ferroelectric structures for moderate (of the order of 1 V) gate voltages. The existence of a hysteresis (or anti-hysteresis) in the dependence of electrical resistance of graphene channel on the gate voltage facilitates creating bistable systems for different applications. The use of ferroelectric substrates has already enabled developing of robust elements of non-volatile memory of a new generation. These elements operate for more than $10^5$ switching cycles and store information for more than $10^3$ s. Such systems can be characterised theoretically by ultrafast switching rates ($\sim 10$–$100$ fs). A theoretical analysis has also demonstrated that the structures 'graphene-on-PZT ferroelectric substrate' would result in developing efficient and fast small-sized modulators of mid-IR and near-IR radiations for different optoelectronic applications.

## 1. Introduction

Although graphene has first been obtained as lately as in 2004, an interdisciplinary 'graphene physics' has already formed on the borders of solid state physics, high-energy physics, physical chemistry, and engineering. Today it acquires the features of a new independent discipline characterised with a steady progress [1–3]. The number of scientific articles on the graphene physics has been increasing dramatically, and the pioneering work by A. Geim and K. Novoselov [4], where electric field effect in single-atomic carbon layer placed on $SiO_2$ substrate is measured for the first time, has been cited for almost 7200 times in January 2012. The Nobel prize awarded to Geim and Novoselov in 2010 has given a great impact to the physics of graphene.

Ukrainian theorists have contributed notably to development of the graphene physics. In particular, V. Gusynin and S. Sharapov (M. Boholyubov Institute of Theoretical Physics, Kyiv, Ukraine) have predicted unconventional integer quantum Hall effect [5, 6], a shift in the quantum magnetic oscillations [7], including Shubnikov–de Haas oscillations [8], and a concentration dependence of cyclotron mass [8, 9]. It has been revealed that the quantum Hall effect for Dirac fermions in the graphene has uncommon character, being characterised with the occupation factor $v = \pm 4(n + 1/2)$, $n = 0, 1, \ldots$ . This anomaly of the quantum Hall effect is caused by degeneration of the lowest Landau level, which amounts to half the degeneration degree of higher levels. Experimental observations of this anomaly have proved unambiguously the fact that quasi-particles near the *K*-point in the graphene, where *c*- and *v*-bands touch with each other, are governed by the Dirac-type equation rather than the Schrödinger one, as for the other materials characterised with translational symmetry. A universal optical conductivity and its threshold-like dependence on the carrier concentration [9] have also been observed experimentally, thus giving another evidence for possible applications of optical properties of the graphene in the IR optics and optoelectronics.

Another field where the contribution of Ukrainians has been essential is a non-equilibrium carrier physics, the theme elaborated extensively in the recent years (see, e.g., the review [10]). A central figure in this activity is F. Vasko (V. Lashkaryov Institute of Semiconductor Physics, Kyiv, Ukraine). Together with his co-workers, he has analysed, in the frame of quasi-classical approach, the transport phenomena occurring in both the intrinsic and doped graphenes, which are caused by heating of carriers with a constant electric field and by photo-excitation of electron-hole plasma. These calculations have been based on the general methods of quantum kinetic theory worked out in the work [11]. The achievements of the Ukrainian theorists in the graphene physics have already been described in detail in the reviews [12, 13].

In the initial stage, it had been graphene itself that attracted all the attention of researchers, whereas a substrate and a gate played only secondary roles, enabling 'doping' of the graphene with electrons or holes. An understanding, however, had appeared soon that the graphene can have important practical applications when it is combined with some other components. This is why the studies of graphene in its interaction with substrates, contacts, phonon and photon thermostats, and the other factors that determine peculiarities of the transport phenomena have become increasingly urgent. The properties of graphene placed on the substrates with high dielectric permittivities $\kappa$ have been extensively examined (see, e.g., the works [14, 15]). Really, substitution of traditional quartz by the substrates with high $\kappa$ values (AlN, $Al_2O_3$, $HfO_2$, $ZrO_2$, etc.) allows obtaining higher carrier concentrations for the same gate voltages. Moreover, it has been assumed that the effective screening of the Coulomb field of charged impurities in the substrate bulk and on its surface should decrease scattering of carriers in the graphene by these impurities and, as a consequence, this should increase the carrier mobility.

However, these expectations have not come true. As shown in the study [15], a decrease in the Coulomb scattering by the impurities in the real systems is accompanied by essential enhancement of scattering by the surface phonon modes. Moreover, the highest permittivity one can get for such traditional linear dielectrics is only 24 (for $ZrO_2$). Therefore an idea has come to substitute a linear-dielectric substrate by a ferroelectric one, where the permittivity can be orders of magnitude higher. As a result, utilisation of ferroelectric substrates (organic or those based on $Pb(Zr_xTi_{1-x})O_3$ abbreviated hereafter as PZT), which behave as dielectrics with extremely high permittivities (up to 3850) under low electric voltages, appears to be fruitful when constructing modulators for the near-IR and mid-IR ranges. The latter can be promising for low-voltage devices with optical on-chip interconnections [16]. Furthermore, the existence of hysteresis in the dependence of ferroelectric polarisation on the field applied makes ferroelectrics promising for constructing memory units [17].

This review is devoted to the first works examining the physical properties of graphene placed upon ferroelectric substrates, which have started vigorously since 2009, as well as practical achievements in applications of the 'graphene-on-ferroelectric' structures for constructing non-volatile memory of a new generation and modulators of optical radiation.

## 2. Ferroelectrics: physical properties and applications

Ferroelectrics are characterised by a spontaneous polarisation, direction and magnitude of which can be changed by an external electric field (see, e.g., [18, 19]). The term "ferroelectrics" outline obvious analogy between ferroelectricity and ferromagnetism, although the most widespread ferroelectrics do not contain iron. Ferroelectricity has been discovered in 1920 by the American physicist Joseph Valasek (born in Czech lands of Austro-Hungary), during his studies on a so-called Rochelle salt [20]. This salt had been fabricated by the French druggist Segnet in the city of La Rochelle in the end of the 17$^{th}$ century. In spite of extensive studies of ferroelectrics, the number of citations of the pioneering paper by Valasek published in the 'Physical Review' [20] is relatively low (95 in February 2012). In other words, the phenomenon has been studied in a versatile manner, though its discoverer remains almost forgotten.

Typical dependences of the electric polarisation on the external field for dielectrics, paraelectrics and ferroelectrics are presented in Fig. 1. The polarisation of a dielectric is linear, while the permittivity of a paraelectric depends on the electric field. The field dependence of the ferroelectric polarisation has a hysteretic shape, similar to the known hysteresis appearing in the ferromagnetic magnetisation. The ferroelectrics reveal such a hysteretic behaviour only below a Curie temperature $T_c$, at which a phase transition occurs. At higher temperatures ferroelectrics behave as paraelectrics. Phase transitions into ferroelectric phases are often described in terms of charge displacement (e.g., in $BaTiO_3$) or 'ordering-disordering' (e.g., in $NaNO_2$), in spite of the fact that a mixed type of behaviour is typical for the majority of ferroelectrics.

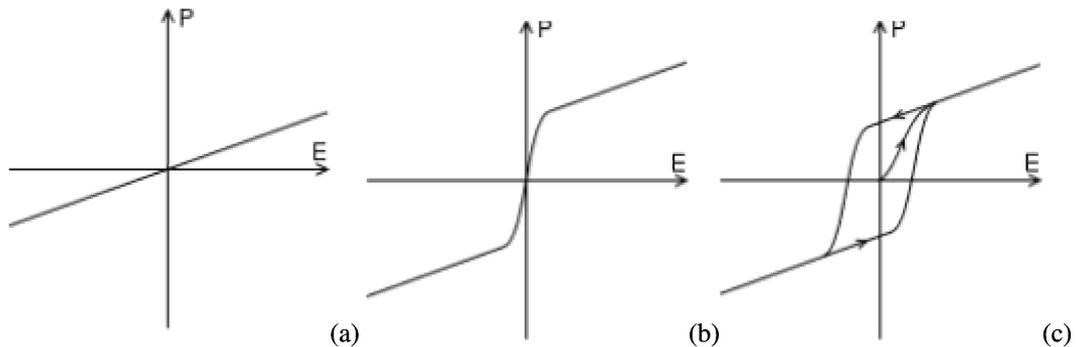

**Fig. 1.** Polarisation *P* as a function of external electric field *E* for the dielectrics (a), paraelectrics (b) and ferroelectrics (c).

A ferroelectric compound $Pb(Zr_xTi_{1-x})O_3$ (PZT) is important for many practical applications. This ceramic of a perovskite type has been obtained for the first time by the researchers from Tokyo Technological Institute (Japan) in 1952. It is characterised by rather high Curie temperature (of the order of 400 °C ), which is dependent on the composition parameter $x$. Notice that one of the PZT components ($PbTiO_3$) is ferroelectric and the other ($PbZrO_3$) antiferroelectric. Consequently, variation of the composition $x$ enables one to obtain ferroelectrics with a wide range of properties. PZT's with small $x$ are often used in different memory devices, while those with $x$ close to 0.52 remain preferable for piezoelectric applications (e.g., ultrasonic transformers and actuators), because the piezoelectric constants of the latter compound have a singularity and their permittivity can be extremely high at the morphotropic phase boundaries [21].

The PZT ceramic is used both as a bulk crystal and a thin film obtained by a chemical vapour deposition (CVD) method. Such a technology is well elaborated for the $Pb(Zr_{0.3}Ti_{0.7})O_3$ compound called also as PZT 30/70. The properties of this alloy can be notably modified by doping with lanthanum, resulting in the PLZT alloy $Pb_{0.83}La_{0.17}(Zr_{0.3}Ti_{0.7})_{0.9575}O_3$ (or PLZT 17/30/70) [22]. The permittivity of the PZT falls in the region 30–3850, depending on composition of the ceramic, its doping, and orientation.

Capacitors with controllable capacitance have been constructed on the basis of ferroelectrics long ago. Their construction is very simple: they consist of a ferroelectric placed between two metallic electrodes. Because of high ferroelectric permittivities, these capacitors are much smaller, when compared with the ordinary ones having the same capacitance, and they are used in a ferroelectric random access memory for computers [23]. Moreover, thin ferroelectric films with the thicknesses of 10–100 nm can provide fields enough for reversing polarisation direction in low-voltage devices. Unfortunately, a quality of the materials, interfaces and the contacts remains crucial for proper operation of the ferroelectric random access memory [24].

The other physical principles used when fabricating non-volatile random access memory based on the nanoscale ferroelectric films have been analysed in the review [25]. It should be noted that miniaturisation of such devices meets with principal limitations caused by the fact that the ferroelectricity is a collective phenomenon and so the ferroelectric properties of any system depend on its size at nanoscale. For instance, a ferroelectric can transform into antiferroelectric or paraelectric its size becomes less than a critical size (for more details see the review [26]).

A combination of possible memory, piezoelectric and pyroelectric characteristics make ferroelectrics very attractive as sensor materials. The ferroelectric capacitors with controllable capacitance are used in medicine (especially in ultrasonic diagnostics, where they generate ultrasonic signals and later detect their echoes), high-quality IR cameras (e.g., a 2D system of ferroelectric capacitors can detect the temperature differences as small as $10^{-6}$ K), sensors of fire danger, vibration sensors, fuel injectors for diesel engines, etc.

In the recent years, a ferroelectric tunnel transition of electrons through thin (several nanometres thick) ferroelectric films placed between two metallic electrodes has been extensively studied [27, 28]. At first it has been predicted theoretically that the piezoelectric effects occurring on interfaces, together with external depolarisation field, would cause a giant electrical resistance effect and an asymmetry in the voltage-current characteristics, which can be used for switching between the states '0' and '1' in logical units. A possibility for practical observations of this effect had been doubted until a principal feasibility of ferroelectric polarisation in ultrathin films with the thicknesses of 3–5 lattice constants was demonstrated experimentally (see, e.g., [25]).

A new class of multiferroics, with both ferroelectric and ferromagnetic orderings, has also been extensively explored. Multiferroics such as $BiFeO_3$, $La_{1-x}Sr_xMnO_3$ and $La_{1-x}Sr_xAlO_3$, are extremely promising for many technological applications, especially in spintronics (see the review [29]).

In 1979, Sven Torbjörn Lagerwall and Noel Clark have produced liquid ferroelectrics based on introduction of chiral impurities into non-chiral smectic liquid-crystalline matrices. These systems easily pass from one stable state to another with the electric field direction being switched [30]. The liquid ferroelectrics manifest the switching times $10^3$ times smaller than those typical for the nematics used is ordinary displays. Flat screens have been constructed on the basis of the Lagerwall–Clark effect and their mass production has been organised by the "Canon" in 1994.

# 3. Graphene on ferroelectric substrate: transport characteristics and non-volatile memory structures

As mentioned before, the main shortage of traditional quartz (SiO$_2$) substrates is their comparatively low permittivity ($\kappa = 3.9$), which limits carrier concentrations in the graphene. The concentration $n$ of carriers in the gated graphene depends linearly upon the gate voltage $V_g$ and the permittivity $\kappa$ of the substrate, being inversely proportional to the substrate thickness $d$:

$$n(\text{cm}^{-2}) = 7.2 \times 10^{10} \frac{300}{d(\text{nm})} \frac{\kappa}{3.9} V_g(V). \tag{1}$$

The figures appearing in Eq. (1) are normalised to the characteristics of SiO$_2$ substrate with the thickness of 300 nm, which has been used in the first works on the graphene problems (see [1–4]). Hence, the highest values of the concentration (and the conductivity) are determined by the substrate breakdown field, which amounts to about 0.5 V/nm for the SiO$_2$ substrate.

Making use of ferroelectric substrates theoretically allows reaching the carrier concentrations in the graphene as high as $6 \cdot 10^{14}$ cm$^{-2}$, i.e. $10^2$ times higher than that peculiar for the traditional SiO$_2$ substrate. This corresponds to the Fermi energy of the order of 1 eV, so that the band spectrum is no longer linear for such energies. Moreover, ferroelectric substrates that reveal spontaneous electric polarisation facilitate creating non-volatile memory units. The first device of this kind has been constructed by the group from Singapore (see the work [17]). The states '0' of '1' of a binary system have been maintained respectively by a high and low resistances of a graphene conducting channel in a field effect transistor. Switching between these two states occurs due to changes in the thin-film ferroelectric polarisation occurring under variations of the gate voltage. These hybrid graphene–ferroelectric devices support a non-volatile resistance variation of about 200 per cent.

A scheme of device implemented in the work [17] is presented in Fig. 2a. A graphene layer is placed onto a traditional quartz substrate, which is put upon a Si gate. A layer of a liquid ferroelectric PVDF-TrFE (approximately 0.7 μm thick) is placed over a graphene sheet and a top Au gate is placed above it (notice that Fig. 2b represents a Hall contact geometry). Continuity of the ferroelectric film has been checked with an atomic force microscope (see Fig. 2d). The characteristics of such a 'graphene-on-quartz' structure have been examined before the liquid ferroelectric is placed onto it. The dependence of the graphene channel resistance on the Si back gate voltage $V_{BG}$ has appeared to reveal a broad maximum (see Fig. 2c). The electron mobility ~ 1500 cm$^2$V$^{-1}$s$^{-1}$ has been determined from the slope of the $R(V_{BG})$ curve.

For all of the 15 samples under test, the dependence of the resistance upon the Au top gate voltage $V_{TG}$ has the shape depicted in Fig. 3a. This dependence displays a completely symmetric hysteretic loop for the top gate voltage sweeps ranging from 0 to +85 V, then from +85 V to –85 V, and finally from –85 V back to 0. The relative changes in the resistance $\Delta R / R = (R_{max} - R_{min}) / R_{min}$ are as large as 3.5 for some of the samples.

This $R(V_{TG})$ dependence is governed by the peculiarities of dependence of ferroelectric polarisation on the field applied (see Fig. 1c). The carrier concentration in the graphene results from both the field caused by the top gate voltage and the field of dipoles on the graphene–ferroelectric interface. Therefore the graphene can remain of *n*-type (or *p*-type) even at the negative (or positive) gate voltages, whenever the latter are not high enough to re-polarise the dipoles of the ferroelectric film. The authors of Ref. [17] have defined the state with the maximum resistance as '1' and that with the minimal resistance as '0'. Fig. 3 shows possible paths of switching between these two states.

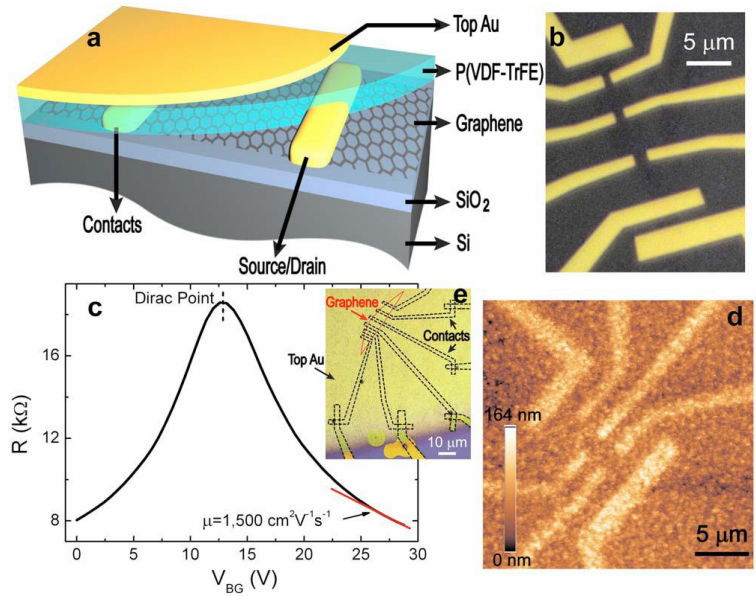

**Fig. 2.** Scheme and characteristics of a non-volatile memory device based on the graphene-on-ferroelectric [17] (see explanations in the text).

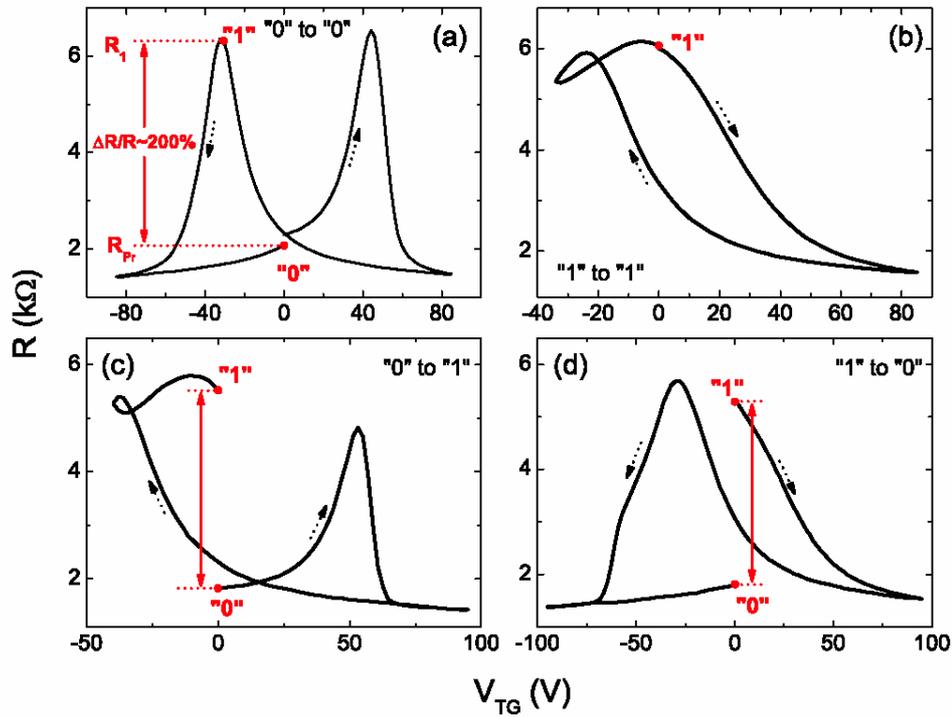

**Fig. 3.** Dependences of graphene channel resistance $R$ on the top gate voltage $V_{TG}$ and paths of switching between the states '0' and '1' in a binary system (see explanations in the text).

Potentially, the hybrid memory on the basis of graphene-on-ferroelectric substrate should be characterised by extremely high switching rates. The graphene-on-ferroelectric channel can have the upper theoretical limit of carrier mobility as high as 200000 cm$^2$V$^{-1}$s$^{-1}$. The switching time can be less than 10$^{-13}$ s for the switching voltages of the order of 1 V and the channel lengths of the order of microns. In other words, this device seems to be a promising candidate for the principal element of super-fast non-volatile memory of a new generation.

It their subsequent work [31], the research group from Singapore has achieved essentially higher mobility values (~ 4200 cm$^2$V$^{-1}$s$^{-1}$) and relative resistance changes (~ 500 per cent), while reproducing the non-volatile memory during 10$^5$ switching cycles for the geometry depicted in Fig. 2.

A similar system with the back and top gates and a 200-nm film of organic P(VDF-TrFE) ferroelectric has been studied very recently in the work [32]. A research team from Switzerland has demonstrated high retention performance for the both memory states, with fully saturated temporal dependence of the graphene channel resistance. This behaviour is distinct from that of 'ferroelectric–polymer–gated silicon' field effect transistors, where the gap between the two memory states continuously decreases in the course of time. Before reaching saturation, the current decays exponentially as predicted by the retention model based on the charge injection into the interface-adjacent layer. The drain current saturation attests to a high quality of the graphene–ferroelectric interface which has a low density of charge traps.

In the study [33] the authors from the United States have reported construction of a field effect transistor on an *n*-layer graphene (*n* = 2–15) placed at a thin film of Pb(Zr$_{0.2}$Ti$_{0.8}$)O$_3$. Fig. 4 shows the resistance of the 7-layer graphene as a function of gate voltage at 300 K. The ferroelectric field permittivity determined on the basis of Eq. (1) and the Hall measurements of concentration is equal to $\kappa$ ~ 100. The dependence $\rho(V_g)$ have had definitely different shapes for the low and high voltages. For the range of sweeps $|V_g|$ < 2 V (see left curve in Fig. 4) the carrier concentration and the resistance are governed by Eq. (1), so that the sweeps from the positive voltages to negative ones are reversible. The Hall mobility obtained for this regime is ~ 70000 cm$^2$V$^{-1}$s$^{-1}$. A small shift of the maximum with respect to zero voltage (here the electro-neutrality point corresponds to $V_g$ = 0.17 V) has been explained by a residual spontaneous polarisation of the ferroelectric film.

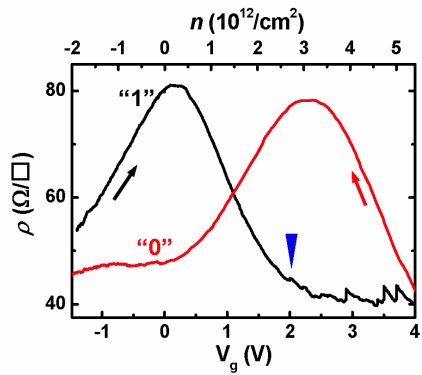

**Fig. 4.** Resistance $\rho$ of a 7-layer graphene channel placed on the Pb(Zr$_{0.2}$Ti$_{0.8}$)O$_3$ film as a function of gate voltage $V_g$ [33].

For the range of sweeps $V_g$ > 2 V, the dependence of the resistance on the gate voltage has a hysteretic shape somewhat similar to that observed in the work [17]. The resistance changes in a manner similar to the right curve in Fig. 4 for the backward voltage sweep (from the maximum value to the zero). The point where the resistance saturates in the forward sweep corresponds to the electro-neutrality point for the backward sweep, and vice versa.

The microscopic mechanism of the effect observed had still remained rather unclear, because the fields in the ferroelectric films that correspond to the gate voltages applied have been much smaller than the coercive fields, which could have reversed the ferroelectric polarisation. Moreover, the hysteresis direction itself has been opposite to that predicted basing on the dependence of ferroelectric polarisation on the external field. This behaviour (reproduced later in all the works on the graphene-on-PZT substrates) has been called in the work [33] as an 'anti-hysteretic' one.

In spite of incomprehensibility mentioned above, the anti-hysteresis has turned out to be quite reproducible, being characterised by a long relaxation time. The curves with the left maximum in Fig. 4 have proved to be more stable in the range of low gate voltages, whereas the curves with the right maximum – in the range of upper voltages. After a slow voltage sweep from 0 to 2V, the resistance changes according to the 'left' curve. However, as soon as the voltage is fixed at this point, the resistance slowly increases up to value that corresponds to the curve with the 'right' maximum. This relaxation occurs in an exponential manner, with the time constant $\tau$ equal to 6 hours for 300 K and 80 days for 77 K. Treating this process as a thermal relaxation between the two metastable states, the authors have estimated the activation energy $\Delta E$ following from the Arrhenius formula,

$$\frac{1}{\tau} \sim \exp\left[-\frac{\Delta E}{kT}\right]. \tag{2}$$

The value of the energy has fallen within the region of 50–110 meV.

It has been noted [33] that an effect similar to the anti-hysteresis is observed in some other ferroelectric structures (e.g., in carbon nanotubes or epitaxial $BaTiO_3$ films). Probably, then the water molecules associated with the interface play an essential role [34]. A balance of dissociation and recombination of the water molecules ($H_2O \leftrightarrow H^+ + HO^-$) depends on the surface geometry and the external fields, and the anion $HO^-$ trapped by Pb sublattice of ultrathin PZT layers can screen the polarisation. The authors of the work [33] have also suggested utilising the anti-hysteresis effect when constructing memory units. The correlation relative difference $\Delta R/R = (R_{max} - R_{min})/R_{min}$ of the minimal and maximal resistances (the states '1' and '0' – see Fig. 4) is inside the region 2.0–3.5, being dependent on the number of graphene layers, while the high mobility values obtained make high switching rates quite possible.

The combined team from Singapore and Korea [35] has rejected a configuration with the two gates and studied the field effect transistor with single-layer and bi-layer graphenes. The latter have been fabricated, using a chemical vapour deposition (CVD) technique, on Cu placed upon 360 nm thick $Pb(Zr_{0.3}Ti_{0.7})O_3$ substrate. Here a high permittivity ($\kappa \approx 400$) enables carrier concentrations of the order of $10^{13}$ cm$^{-2}$ for the gate voltages ~1 V. The relationships $n = \alpha V_g$ ($\alpha = 6.1 \times 10^{12}$ cm$^{-2}$V$^{-1}$) is valid inside the linear regime.

The ferroelectric polarisation causes essential hysteresis in the dependence of resistance on the gate voltage beyond the linear region (at $V_g > 1.1$ V – see Fig. 5). The increase in the ferroelectric polarisation leads to increasing distance between the two maxima. The main result of the work [35] is that the non-volatile memory can be successfully fabricated basing on cheap CVD-graphene, rather than more expensive on-scotch-type 'exfoliated' graphene.

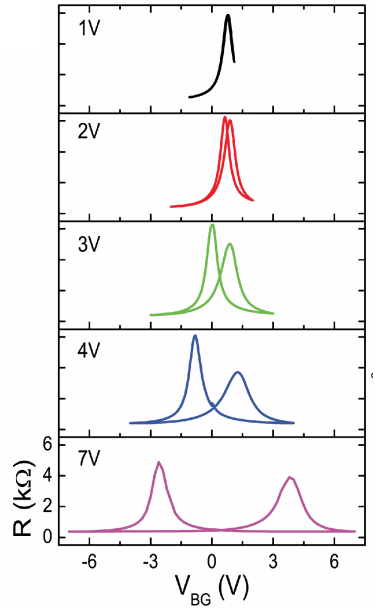

**Fig. 5.** Dependences of resistance $R$ of the graphene-on-PZT 30/70 substrate upon gate voltage $V_{BG}$ for different sweep voltages shown in the legends [35].

The research group from the United States and Korea [36] has reported fabrication of a robust non-volatile memory based on the single-layer graphene placed on the ferroelectric PZT substrate ($\kappa \sim 400$–500). The scheme of their device is sketched in Fig. 6d. A careful optical and Raman characterisations of the system has been carried out. It has been shown (see Fig. 6a, b) that the three-layer system comprising a graphene, a ferroelectric, and a Pt gate, reveals the highest contrast at the PZT-layer thickness of 180 nm, and so the graphene can become visible in the region of visible light. A presence of distinct 'G' and '2D' peaks in the Raman spectra (see Fig. 6c) demonstrates a high quality of single-layer graphene produced by both the exfoliation and the CVD methods.

The electrical measurements have been performed in vacuum ($1.1 \times 10^{-6}$ Torr) at a constant source-drain voltage $V_{ds}$. The dependence of the current $I_d$ in the source-drain circuit on the gate voltage $V_g$ for the exfoliated graphene has a characteristic shape with two minima for the sweep voltages $V_{g(sweep)} > 1$ V (see Fig. 7a). These two conductivity minima correspond to the voltages $V^+_{g\,min}$ and $V^-_{g\,min}$, respectively. The Fermi level crosses the Dirac point at such voltages, and the electrostatic potential of dipoles in the ferroelectric is balanced by the potential of interface-absorbed impurities. The minimal conductivity value at low temperatures is determined by a well-known quantum limit $4e^2/\hbar$ [3]. The asymmetry in the electron and hole conductivities has been explained in the work [36] by doping of the graphene with metals near the interface and by metallic contacts of the source and the drain.

A similar dependence for the CVD-produced graphene is displayed in Fig. 7b. A difference from the corresponding dependence for the exfoliated graphene has been explained by essential chemical doping associated with etching of a Cu sheet, onto which the graphene is primarily deposited. This chemical doping lowers the Fermi level down the Dirac point and suppresses elec-

Ukr. J. Phys. Opt., V. 13, Suppl. 3 Sci. Horiz.                                                                          S23

tronic conductivity. Despite this fact, the dependence has also a hysteretic form, with two definitely different states of high and low conductivities.

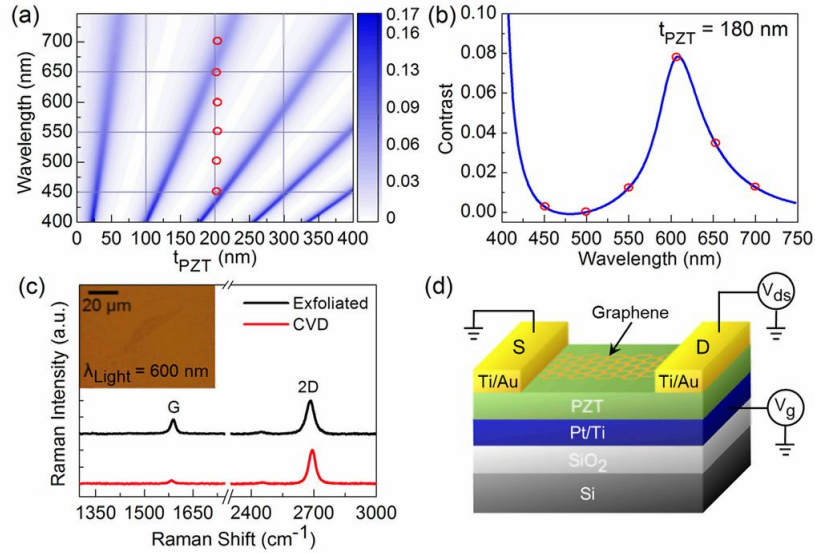

Fig. 6. (a) Map of contrast for the system 'graphene–ferroelectric–gate' for different light wavelengths and PZT-substrate thicknesses $t_{PZT}$, (b) contrast as a function of wavelength for the substrate thickness of 180 nm, (c) Raman spectra of exfoliated (upper curve) and CVD-based (lower curve) graphenes, and (d) scheme of a memory unit [36].

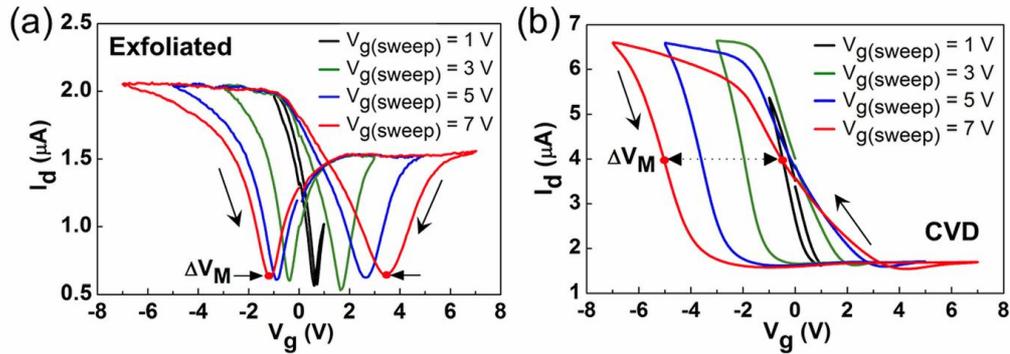

**Fig. 7.** Dependences of current $I_d$ in the source-drain circuit on gate voltage $V_g$ for the exfoliated (a) and CVD-produced (b) graphenes [36].

The upper theoretical limit of the 'memory window' $\Delta V_M$ (i.e., the width of the hysteresis loop) is determined by the voltage $V_C$, which corresponds to the ferroelectric coercive force ($\Delta V_M = 2V_C$). However, this 'memory window' has been introduced in the work [36] in somewhat different way as shown in Fig. 7. Namely, the 'memory window' is given by the interval between the two crossings of the Dirac point by the Fermi level in the exfoliated graphene, and it is equal to the width of the hysteresis loop for the CVD-produced graphene, which is always of *p*-type. As one can clearly see, the 'memory window' reaches its saturation approximately at 7 V with increasing sweep voltage. This value is essentially smaller for ordinary semiconductor field effect transistors. The two states '0' and '1' can be stored in the system under examination for about $10^3$ s.

## 4. Mechanism of anti-hysteresis in the resistance of graphene-on-PZT

The qualitative mechanism of the anti-hysteresis observed in the dependences $I_d(V_g)$ for all the graphene-on-PZT systems has been suggested in the study [36]. It essentially involves the interface states (see Fig. 8). It has been supposed that, for the forward gate voltage sweep at $V_{g1}$, the resultant of the external field $E_{ex}$ and the field $E_p$ of the ferroelectric dipoles dopes graphene with electrons (see Fig. 8a). The electrons are trapped by the interface states, with a further increase in the gate voltage (see Fig. 8b). The electrons remain on the interface states for the backward sweep, thus lowering the electrochemical potential of the PZT by some value $\Delta\mu$ (see Fig. 8c) and reducing the field $E_p$. When the gate voltage reaches the value $V_{g1}$ on the backward sweep, the electron concentration would be lower than the concentration for the same gate voltage in its forward sweep (see Fig. 8d).

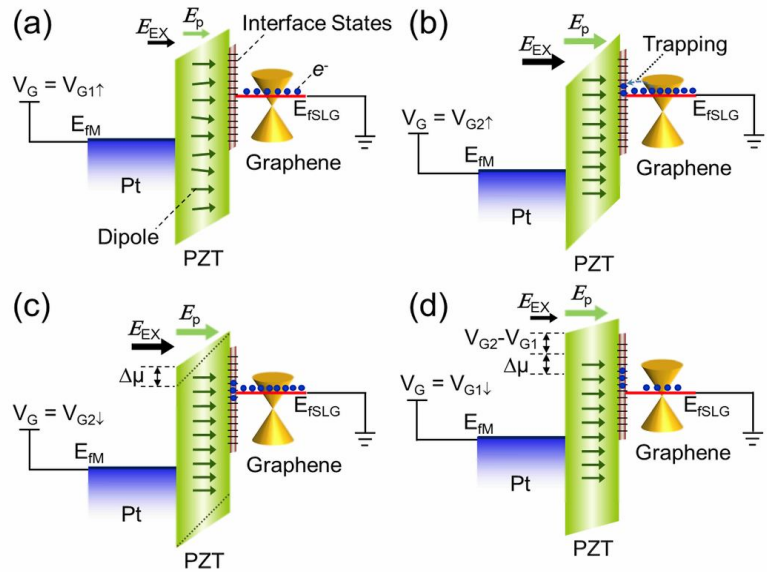

**Fig. 8.** Illustration of physical mechanism of 'anti-hysteretic' $I_d(V_g)$ behaviour in the graphene-on-PZT substrate, which has been suggested in the work [36].

A quantitative model of anti-hysteretic behaviour of the resistance of the graphene-on-PZT substrate has been developed in the work [37]. The gated single-layer graphene has been examined there, assuming that the Fermi energy depends on the concentration as

$$E_F = \hbar v_F (\pi n)^{1/2}, \qquad (3)$$

where $v_F = 10^8$ cm/s. It has been supposed that some interface state exists, with the corresponding energy $E_T$. On the forward $V_g$ sweep (when $E_F < E_T$), the carrier concentration is governed by a simple relation

$$n = \kappa V_g / 4\pi e d, \qquad (4)$$

where $d$ denotes the substrate thickness. However, the electrons from the gated graphene are captured by the interface states of a great 2D density $n_T$, whenever $E_F = E_T$. The negative charge of the occupied interface states screens the field in the substrate, so that the concentration of carriers in the gated graphene is given by

$$n = \kappa V_g/4\pi e d - n_T \qquad (5)$$

for the further forward $V_g$ sweep.

The next assumption is that the lifetime of electrons on the interface states is much longer than the switching time of the system. As a result, Eq. (5) is valid for the backward sweep too, and the general dependence of the *n* parameter on the gate voltage has a hysteretic shape displayed in Fig. 9 (see curves 1 and 2, where arrows indicate directions of sweeps). Now curve 2 reaches the Dirac point at some gate voltage $V_{DP}$ determined by the concentration $n_T$ of the interface states as follows:

$$V_{DP} = 4\pi e d n_T/\kappa. \qquad (6)$$

Note that Fig. 9 (curves 1 and 2) presents the holes concentrations left to the Dirac points. The trapped electrons recombine with the holes in the graphene sheet at large negative $V_g$ values and the *n* parameter is again governed by Eq. (4).

The total resistance of the graphene sheet is a reciprocal of the conductivity, i.e.

$$\rho(V_g) \approx 1/(\sigma(V_g) + \sigma_{min}). \qquad (7)$$

Here the first term in the denominator represents the conductivity of the gated graphene, which changes linearly with $V_g$ and *n*, and the second term is the minimal graphene's conductivity at the Dirac point [3]. The dependence $\rho(V_g)$ is seen in Fig. 9, with curves 3 and 4 corresponding to the forward and backward sweeps, respectively. As one can see, it has a hysteretic shape earlier observed experimentally in the studies [33, 35, 36]. The distance between the Dirac points in curves 3 and 4 is determined by the concentration of the interface states via Eq. (6) and it does not depend on $E_T$ in this rough approximation. The substitution of experimental values taken from Ref. [33] into Eq. (6) yields $n_T = 2.7 \cdot 10^{12}$ cm$^{-2}$. This value seems to be reasonable for the interface between the graphene and ferroelectrics, since even much higher concentrations of the surface states have been observed for the ferroelectrics [38]. Finally, relaxation of the anti-hysteresis observed experimentally [33] can be explained by the finite lifetime of electrons in these states.

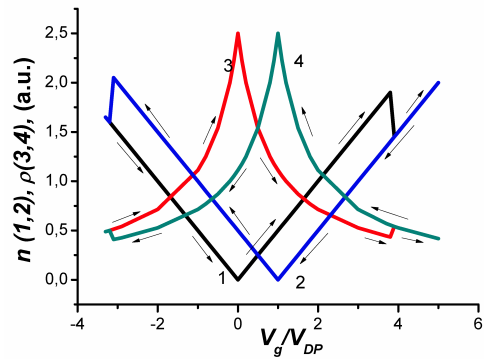

**Fig. 9.** Anti-hysteresis observed in dependences of carrier concentration and resistance on the gate voltage for the graphene-on-PZT: theory [37]. Arrows indicate gate voltage sweep directions.

A simple model presented in the work [37] explains the anti-hysteretic behaviour of the resistance of the graphene-on-PZT by special interface states with great 2D concentrations. (Notice that they can be linked to HO$^-$ anions captured by Pb$^{2-}$ sublattice of the PZT, which create localised states with the binding energy ~ 200 meV [39]; up to the order of magnitude, the latter value corresponds to the thermal activation energy observed in the experiments [33]). The assumptions made above neglect the hysteresis peculiar for the PZT itself. However, this is indeed valid for the region of small $V_g$ which, in its turn, corresponds to the region of $n$ values much smaller than the nominal 2D charge density corresponding to the PZT polarisation ($\sim 3 \times 10^{14}$ cm$^{-2}$) [33].

## 5. Modulation of near-IR and mid-IR radiations using graphene-on-ferroelectric substrate

An essential feature of the optical properties of graphene is its notable interaction with optical radiation in a wide spectral region, from far-IR up to UV, due to effective interband transitions (see [40–42] and references therein). Amplification of optical response of the Fresnel system 'graphene–300 nm thick quartz substrate–Si gate' has allowed for making graphene optically visible already in the first work [4] on the problem. An example of applications of unique optical properties of the graphene is fabrication of graphene-based saturable absorbers for ultrafast lasers in the telecommunication region (see [43]).

Recently, both the modulation [44] and polarisation [45] of the IR radiation have been observed experimentally in the graphene structure integrated with a waveguide. Namely, a graphene-based optical modulator for the near-IR region (1.35–1.6 μm) has been implemented in the work [44]. It has been shown that the modulator can be promising for the devices with on-chip optical interconnections, and its modulation efficiency is already comparable to, if not better than, that typical for traditional semiconductor materials such as Si, GeSi or InGaAs, which are orders of magnitude larger in the active volume. In that work, a single-layer graphene has been placed upon a 7 nm thick Al$_2$O$_3$ substrate over a Si gate, which also serves as a waveguide for the near-IR radiation (see Fig. 10). The footprint of the device is as small as 25 μm$^2$, while the operation speed is as high as 1.2 GHz.

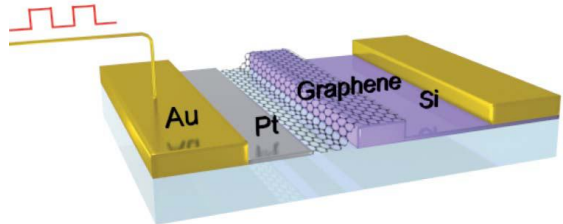

**Fig. 10.** General scheme of device described in the work [44]. A thin sapphire substrate (invisible in the figure) is placed between graphene and silicon.

A general theory of the carrier-induced modulation of radiation with a gated graphene has been worked out in the study [46]. There it has been shown that the contribution of carriers modifies essentially the graphene response due to the Pauli blocking effect. Then the absorption is suppressed at $\hbar\omega/2 < E_F$, where $E_F$ means the Fermi energy. At low temperatures or at high doping levels, the threshold frequency for the absorption jump (when the absorption becomes essential) is specified by the condition (see Fig. 11)

$$\hbar\omega_{th} = 2E_F \sim \sqrt{n} , \qquad (8)$$

where the concentration $n$ is given by Eq. (1).

As already mentioned above, high-$\kappa$ substrates (AlN, Al$_2$O$_3$, HfO$_2$, ZrO$_2$, etc.) enable obtaining higher concentrations for the same gate voltages. This is important because, for the threshold wavelength $\lambda_{th}$ corresponding to the threshold frequency given by Eq. (8), Eqs. (1) and (8) give rise to

$$\lambda_{th} \sim \sqrt{\frac{d}{V_g}} \equiv 1/\sqrt{E_s}, \tag{9}$$

with $E_s$ being the homogeneous gate-induced electric field intensity in the substrate. Eq. (9) implies that, in order to get modulation for smaller radiation wavelengths (e.g., those from the visible range), one needs stronger fields (and higher gate voltages) that eventually can cause a breakdown of the substrate.

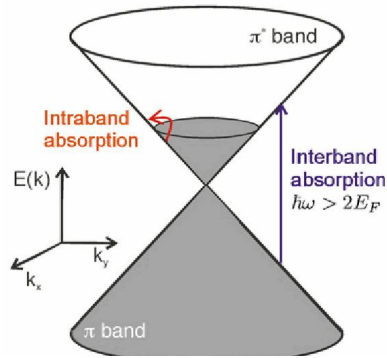

**Fig. 11.** Optical transitions in graphene are permitted for the photon energies $\hbar\omega > 2E_F$.

The transmittance and reflectance coefficients of the 'graphene layer–PZT substrate–Si gate' system may be written as (see [46])

$$T_\lambda = \sqrt{\kappa_{Si}(\lambda)} \frac{|E_t|^2}{E_{in}^2}, \quad R_\lambda = \frac{|E_r|^2}{E_{in}^2}, \tag{10}$$

where $\kappa_{Si}$ is the wavelength-dependent permittivity of the Si gate. The relations among the amplitudes $E$ of the incident (*in*) wave, the wave back-reflected (*r*) into vacuum and the wave transmitted (*t*) through the Si gate may be obtained by solving a system of the wave equations for the vacuum, the substrate and the gate, with proper boundary conditions taking into consideration the absorption due to interband transitions of carriers in the graphene layer.

The calculations performed in the work [46] show that a modulator for the near-IR telecommunication region (~ 1.5 μm) should be based on a single-layer (or multi-layer) graphene placed over a high-$\kappa$ substrate (in fact, this is just the case realised in the study [44]). The calculations mentioned above have been performed for the structures with various degrees of degeneration, under the condition of normal propagation of the wave (see Fig. 12).

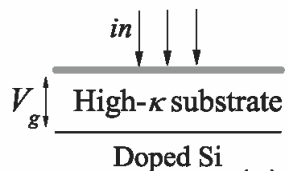

**Fig. 12.** Principal scheme of modulator studied in the work [46]: $V_g$ is the gate voltage.

The results of the work [46] have proved that the gated graphene can efficiently modulate the near-IR radiation in case of high-$\kappa$ substrates and the fields ~ 5 MV/cm. For the case of low-$\kappa$ SiO$_2$ substrate, the fields applied should be essentially stronger (~ 20 MV/cm), which is comparable to the breakdown value. The structure 'graphene-on-quartz' can nevertheless efficiently modulate the mid-IR radiation (see Fig. 13)

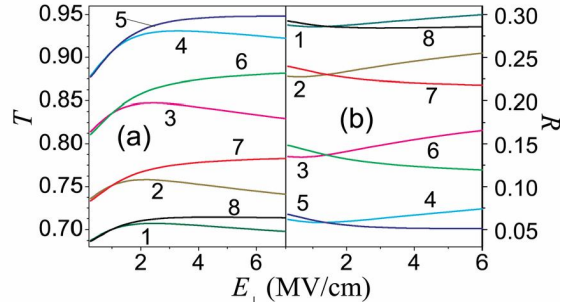

**Fig. 13.** Reflectance ($R$) and transmittance ($T$) of the graphene-on-quartz substrate as functions of in-substrate field ($\lambda$ = 10.6 μm). Curves 1 to 8 correspond to substrate thicknesses 0.3, 0.6, 0.9, 1.2, 1.5, 1.8, 2.1 and 2.4 μm, respectively [46].

The highest $E_s$ values reached in the experiments [44] have been in fact of the order of 5 MV/cm. This has needed, however, an extremely high accuracy of substrate preparation (a 7 nm thick $Al_2O_3$ substrate has been deposited over a Si gate, which also serves as a waveguide, using the atomic layer deposition technique).

Ferroelectric substrates with extremely high permittivities can be fruitful for further development of the modulators based upon the gated graphene. It is important that the epitaxial thin ferroelectric PZT films behave as high-$\kappa$ dielectrics at low voltages ($V < V_{cr} \sim 1$–2 V), with $\kappa = 73$ for x = 0.2 [33] and $\kappa = 400$ for x = 0.3 [35]. This is why they can be used in low-voltage mid-IR modulators based on the gated graphene.

The critical field for the PZT substrate, under which the substrate still behaves as a high-$\kappa$ dielectric, may be obtained from the results [33] ($d$ = 300 nm and $V_{cr} \sim 2$ V) and [35] ($d$ = 360 nm, $V_{cr} \sim 1$ V). This yields $E_{cr}$ = 67 kV/cm for x = 0.2 and $E_{cr}$ = 28 kV/cm for x = 0.3.

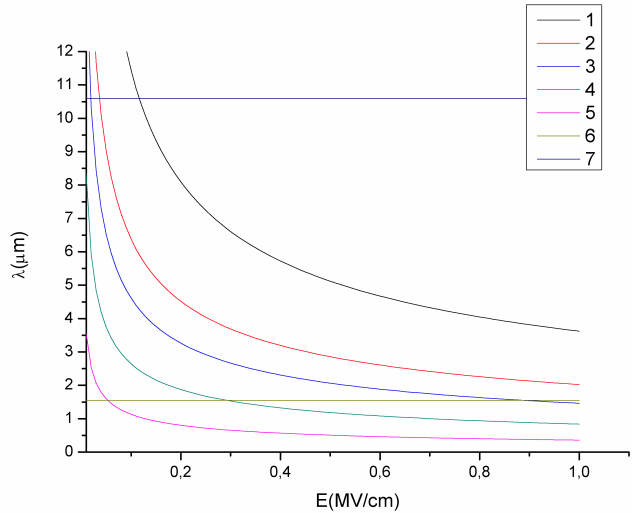

**Fig. 14.** Dependences of threshold wavelength (see Eq. (9)) on field in the substrate for different substrates: curve 1 – $SiO_2$, 2 – $Al_2O_3$, 3 – $ZrO_2$, 4 – PZT (x = 0.2), and 5 – PZT (x = 0.3). Lines 6 and 7 correspond to 1.55 and 10.6 μm, respectively.

Fig. 14 shows dependence of the threshold wavelength (see Eq. (9)) on the in-substrate field for the substrates with different permittivities: $SiO_2$ ($\kappa = 3.9$, curve 1), $Al_2O_3$ (12.53, curve 2), $ZrO_2$ (24.0, curve 3), PZT 20/80 (73, curve 4), and PZT 30/70 (400, curve 5). As seen from Fig. 14, the fields for the PZT that correspond to $\lambda_{th}$ in the telecommunication region ($\lambda = 1.55$ μm, horizontal line) are several times higher than $E_{cr}$. However, the modulation in the mid-IR range (namely, for the wavelength $\lambda = 10.6$ μm of $CO_2$ lasers) can be accomplished for the fields essentially lower than $E_{cr}$, when the PZT behaves as a high-$\kappa$ dielectric that reveals extremely high permittivities.

The reflectance and transmittance coefficients calculated according to Eq. (10) are shown for both the single-layer (Fig. 15a and Fig. 16a) and 5-layer (Fig. 15b and Fig. 16b) graphenes placed on the PZT substrates with x = 0.3. Here different curves correspond to the films of different thicknesses (240, 280, 320, 360 and 400 nm). The dielectric permittivity of the PZT at $\lambda = 10.6$ μm is taken to be equal to 5 (see [16]). As one can see, the absorption jump leads to essential jump in the reflection and the transmission. Here the fields are approximately equal to 2.5–3 kV/cm for the single-layer graphene and 13–15 kV/cm for the 5-layer graphene placed on the PZT (x = 0.3; the permittivity $\kappa \approx 400$). The modulation depth can be of the order of 20 per cent (~ 2 per cent for each layer) for the fields much lower than the critical ones that give rise to ferroelectric hysteretic phenomena.

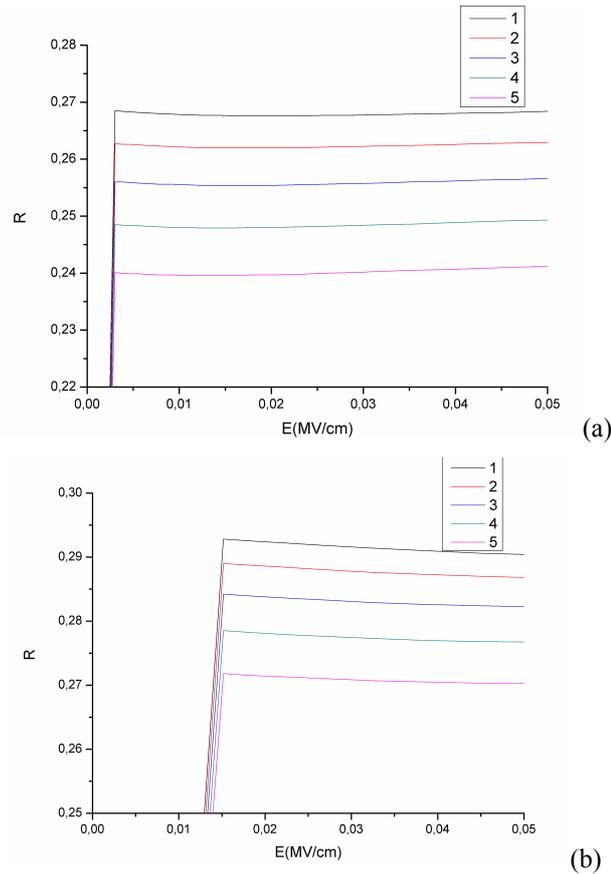

**Fig. 15.** Reflectance coefficients *R* as functions of in-substrate field for the wavelength of 10.6 μm, the room temperature and different substrate thicknesses (curve 1 – 240, 2 – 280, 3 – 320, 4 – 360, and 5 – 400 nm). Panels (a) and (b) correspond to single-layer and 5-layer graphenes, respectively [16].

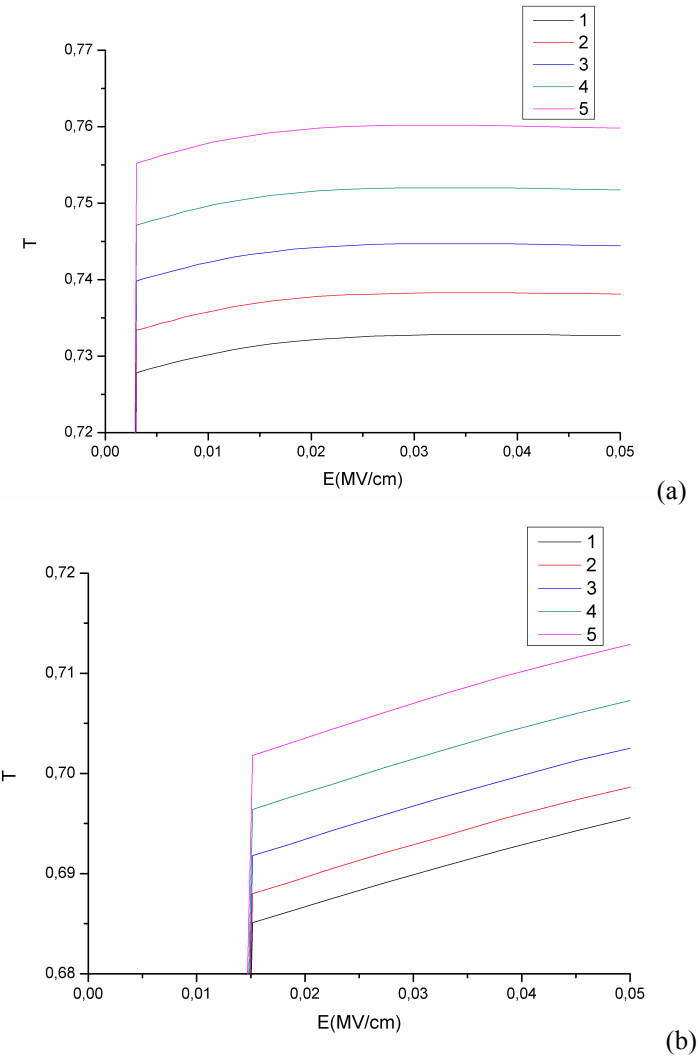

**Fig. 16.** Transmittance coefficients *T* as functions of in-substrate field for the wavelength of 10.6 μm, the room temperature and different substrate thicknesses (curve 1 – 240, 2 – 280, 3 – 320, 4 – 360, and 5 – 400 nm). Panels (a) and (b) correspond to single-layer and 5-layer graphenes, respectively [16].

A comparison of Fig. 15 and Fig. 16 with Fig. 13 demonstrates that the electric fields in the substrate, under which efficient modulation of the mid-IR radiation occurs, are two orders of magnitude smaller for the graphene-on-ferroelectric substrate than those for the system graphene-on-quartz. Moreover, the 'modulation edge' is much more acute in the first case. This is of great significance for the devices operating at low switching voltages.

The results described above demonstrate practical possibilities for fabrication of low-voltage modulators for the mid-IR range basing on the gated graphene placed upon the ferroelectric PZT substrates. The advantage of such modulators, when compared with the device built in the work [44] using atomically deposited 7 nm thick $Al_2O_3$ substrate, is simplicity of preparation of the epitaxial PZT film substrates. The modulators can potentially operate at 500 GHz, because the typical times of the speed-limiting processes such as carrier recombination and generation in the graphene are of the order of picoseconds. Moreover, as the modulation of the mid-IR radiation remains efficient enough not only for the transmittance but also for the reflectance parameters (see Fig. 15), the geometry of the modulator may be much simpler than that shown in Fig. 10 (with the gate playing a role of a waveguide).

In principle, the modulation of the near-IR light can be based on the same mechanism, though the PZT substrates should then reveal higher $\kappa$ values. It is known that the PZT features extremely large $\kappa$'s at the morphotropic phase boundary (near $x = 0.52$) [21]. Then the dielectric constant of the PZT can be as high as 3850, depending additionally upon the orientation and doping. The authors of Ref. [35] have measured experimentally the values as high as $\kappa = 2000$, using the doping that consists in partial substitution of Pb by La, and finely tuning the ratio of Zr and Ti contents. Such the permittivity values should be quite sufficient for the near-IR modulation.

## 6. Bistable optical system based on hysteresis observed in the reflectivity of graphene-on-PZT system

It has been predicted theoretically in the work [47] that the anti-hysteretic behaviour of the concentration in graphene-on-PZT, which has been observed experimentally in Refs. [33, 35, 36] and explained in Ref. [37] by the screening of electric field in the substrate by some interface levels, can also reveal itself in optics. The reflection of the 'graphene layer–PZT substrate–Si gate' system under normal propagation of light with the wavelength $\lambda$ has also been studied there. Notice that the threshold wavelength $\lambda_{th}$ given by Eq. (9), which corresponds to the threshold frequency for which the modulation begins (see Eq. (8)), may be written with Eq. (3) as

$$\lambda_{th} = \frac{\sqrt{\pi} c}{v_F \sqrt{n}}. \qquad (11)$$

Here $c$ is the light velocity in vacuum. In case of $n = n_T$, the modulation edge in the $V_g$ scale corresponds to the point $V_{DP}$. Then the reflectivity $R$ at this point rapidly increases and reaches the value approximately equal to $0.023N$, where $N$ is the number of graphene layers. With further increase in $V_g$, the parameter $E_F$ becomes equal to $E_T$. Then the electrons are trapped on the interface levels, the electron concentration in the graphene is governed by Eq. (5), the $E_F$ value decreases and can become smaller than that corresponding to $V_{DP}$ (see Eqs. (3) and (4)). This means that the graphene layer can no longer modulate the radiation with the wavelength $\lambda$, because the direct interband transitions are now forbidden due to the Pauli blocking effect. Therefore the reflectivity $R$ decreases down to its initial level (see curve 3 in Fig. 17).

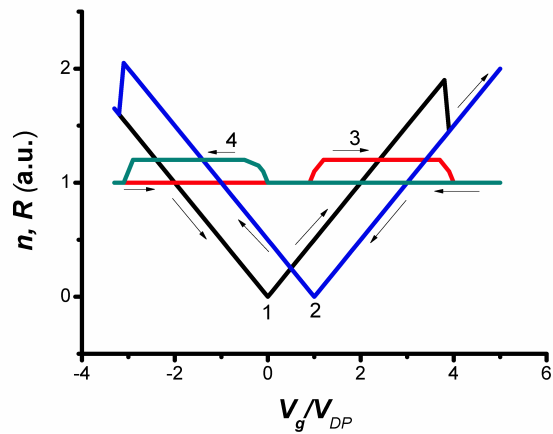

**Fig. 17.** Anti-hysteresis effects observed for carrier concentration (curves 1 and 2) and reflectivity (curves 3 and 4) of the graphene-on-PZT system: theory [47]. Arrows indicate gate voltage sweep directions.

For the backward $V_g$ sweep, the concentration $n$ follows curve 2 of Fig. 17 and the reflectivity $R$ – curve 4 of that figure. This implies that the modulation of radiation with the wavelength $\lambda$ associated with the interband transitions of holes starts now at $V_g = 0$ (each of the curves 1 and 2 in Fig. 17 is symmetric with respect to the corresponding Dirac point – see Eqs. (1)–(4)). The $R$ parameter increases at this point and decreases back to its initial level when the electrons on the interface states recombine with the holes in the graphene sheet and the concentration of the holes decreases.

This hysteresis in the $R$ parameter can be observed experimentally and applied in fast bistable systems. The latter are suitable for non-volatile memory devices with on-chip optical interconnections. Here a practical question arises whether the effect can be applied for modulating the radiation from the telecommunication range (e.g., $\lambda = 1.55$ μm widely used with $SiO_2$ fibres). Inserting the latter wavelength into Eq. (11) and taking Eqs. (4) and (5) into account yields in $n_T \approx 1.2 \times 10^{13}$ cm$^{-2}$. This value seems to be quite realistic for the graphene–ferroelectric interfaces [38], since the concentrations of this order of magnitude have been observed experimentally in Refs. [33, 35, 36]. The estimations based on Eqs. (11), (4) and (5) testify that the effect occurs at low gate voltages $V_g = 2$–3 V, which also makes it attractive from the viewpoint of possible applications.

## 7. Conclusions

The system 'graphene-on-ferroelectric' has a number of unique features. The possibilities for obtaining high ($\sim 10^{12}$ cm$^{-2}$) carrier concentrations at the moderate (of the order of 1 V) gate voltages and the existence of hysteresis (or anti-hysteresis) in the dependence of the resistance of graphene channel on the gate voltage are among these features. Utilisation of the ferroelectric substrates for the graphene should enable constructing of robust elements of non-volatile memory of a new generation [17, 31–33, 35, 36]. These elements work for more than $10^5$ switching cycles and keep information for more than $10^3$ s. As a matter of fact, such systems are characterised theoretically by the ultrafast switching rate ($\sim 10$–100 fs).

In order to explain the anti-hysteretic dependence of the resistance of the graphene-on-PZT ferroelectric substrate on the gate voltage, a quantitative model has been suggested in the work [37]. This model, which takes into consideration screening of the electric field in the substrate by the electrons captured by the states associated with the graphene–ferroelectric interface, explains fairly well all the experimental data obtained up to now. Its estimations may be useful while fabricating the non-volatile memory. It is based on bistable systems, where the logical '0' corresponds to one value of the resistance of graphene and '1' to the other.

It has also been proved theoretically that efficient, fast and small-sized modulators of the mid-IR and near-IR radiations can be constructed for different optoelectronic applications, basing on the graphene placed upon the ferroelectric PZT substrates [16]. The modulation depth for the 5-layer graphene can be of the order of 10 per cent in the region of gate voltages where thin epitaxial PZT films behave as high-$\kappa$ dielectrics.

A model for the hysteretic behaviour observed in the reflectivity of the system 'graphene–ferroelectric PZT substrate–gate' under the gate voltage variations has been developed and analysed in the study [47], including the effects of electron trapping at the graphene–PZT interface states. It has been demonstrated that the hysteresis in the reflectivity can be observed experimentally for the telecommunication wavelength $\lambda = 1.55$ μm at low gate voltages. This can be used while creating fast bistable systems for the novel non-volatile memory devices with on-chip optical interconnections.

Beside of the problems discussed above, the experimental results reported in the works [33, 35, 36] are of substantial and general physical value. It has not been clear until now, which number $N$ of layers makes a multi-layer graphene to turn into ultrathin graphite (see, e.g., [48]). In his Noble lecture [49], Andre Geim has insisted that this number is in fact equal to $N = 2$: the bi-graphene represents a unique case that differs from the single-layer graphene in many aspects (a parabolic spectrum near the band edge, a presence of a gap in the electric field, etc.). On the contrary, the results for the anti-hysteresis observed in the resistance of the graphene-on-PZT system under the conditions of gate voltage variations [33, 35, 36] testify very similar behaviours, regardless of the number of graphene layers ($n = 1$–15), the carrier mobility (16000–140000 cm$^2$/Vs), and the dielectric constant of the PZT (30–500). This implies that, at least for some classes of phenomena characterised by rather high carrier energies far from the band edges, both the single-layer and multi-layer graphenes can be treated as the same class of physical objects.

## Acknowledgement

This work was supported by the State Fundamental Research Fund of Ukraine (Grant 40.2/069).

***Анотація.*** *Подано огляд експериментальних і теоретичних робіт за останні три роки, у яких вивчають властивості та можливі застосування графену на сегнетоелектричній підкладці (органічному сегнетоелектрику або кераміці $Pb(Zr_xTi_{1-x})O_3$ (скорочено PZT)). Графен на сегнетоелектричній підкладці має декілька унікальних переваг, порівняно з графеном на підкладці $SiO_2$ або на діелектриках з високою діелектричною проникністю. У легованому затвором графені на сегнетоелектрику можна одержати високі (~ $10^{12}$ см$^{-2}$) концентрації носіїв для невисоких (порядку 1 В) напруг на затворі. Наявність гістерезису (або антигістерезису) на залежності питомого опору графенового каналу від напруги на* *затворі дає змогу створювати бістабільні системи для різних застосувань. Використання сегнетоелектричних підкладок для графену уможливило створення надійних елементів енергонезалежної пам'яті нового покоління. Ці елементи витримують понад $10^5$ перемикань, працюють і зберігають інформацію упродовж понад $10^3$ с. Теоретично їх можна характеризувати ультрашвидким перемиканням (~ 10–100 фс). Теоретично також було доведено, що на основі графену на сегнетоелектричній підкладці PZT можна створити ефективні, швидкодійні й мініатюрні модулятори випромінювання близького та середнього ІЧ-діапазонів.*